\documentclass{aa}
\usepackage{txfonts}
\usepackage{psfig}

\def\ltsima{$\; \buildrel < \over \sim \;$}
\def\simlt{\lower.5ex\hbox{\ltsima}}            

\begin{document}


\title{An extensive library of synthetic spectra covering 
the far red, RAVE and GAIA wavelength ranges}

\author{
       Toma\v{z} Zwitter\inst{1}
\and   Fiorella Castelli\inst{2,3}
\and   Ulisse Munari\inst{4,5}
       }
\offprints{T. Zwitter}

\institute {
University of Ljubljana, Department of Physics, Jadranska 19, 1000 Ljubljana, Slovenia 
\and
CNR-Istituto di Astrofisica Spaziale e Fisica Cosmica, 
Via del Fosso del Cavaliere, 00133, Roma, Italy 
\and
INAF-Osservatorio Astronomico di Trieste,
    Via G.B. Tiepolo 11, 34131, Trieste, Italy
\and
Osservatorio Astronomico di Padova, Sede di Asiago, 36012 Asiago (VI), Italy
\and 
CISAS, Centro Interdipartimentale Studi ed Attivit\`a Spaziali dell'Universit\`a di Padova, Italy
}
\date{Received date..............; accepted date................}

\abstract{
A library of 183\,588 synthetic spectra based on Kurucz's ATLAS9 
models is presented for the far red spectral interval (7653 -- 8747~\AA). 
It is characterized by
3\,500~K $\le T_\mathrm{eff} \le$ 47\,500~K, $0.0 \le \log g \le 5.0$,  
$-3.0 \le [\mathrm{M}/\mathrm{H}] \le +0.5$, 
$0 \le V_\mathrm{rot} \le 500$~km~s$^{-1}$, $\xi = 2$~km~s$^{-1}$.
The whole grid of spectra is calculated for  resolving powers
8\,500, 11\,500 and 20\,000. A section of the grid is
also computed for [$\alpha$/Fe]=+0.4 and for microturbulent velocities  
0 and 4~km~s$^{-1}$.
The library covers the wavelength ranges and resolutions of the
two ambitious spectroscopic surveys by the ground experiment RAVE and 
the space mission GAIA.
Cross-sections across the multi-dimensional data-cube are used to 
illustrate the behaviour of the strongest spectral lines. 
Interpretation of real data will have to include interpolation 
to grid substeps. We present a simple estimate of the accuracy 
of such a procedure.

}
\maketitle

\begin{figure*}[ht]
\centerline{\psfig{file=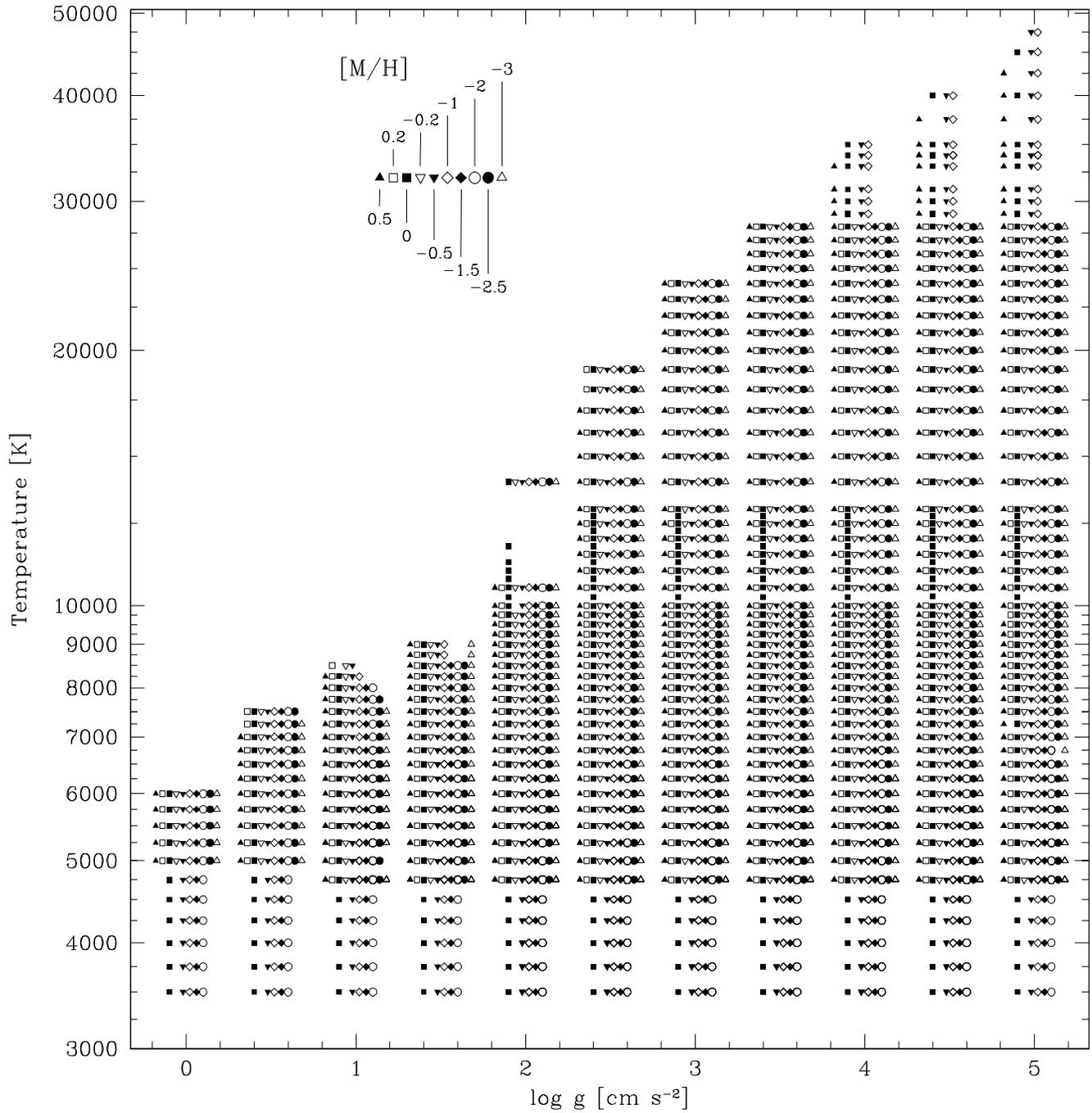,width=20.0cm,height=20.5cm}}
\caption[]{Graph of calculated spectra without alpha enhancement 
and with a micro-turbulent velocity of 2~km~s$^{-1}$. Metallicity is 
coded with different symbols which are plotted with small horizontal 
offsets for clarity. Each spectrum is calculated for 11 ($T_\mathrm{eff}<7000$~K)
or 14 ($T_\mathrm{eff} \ge 7000$~K) different rotational velocities and at three 
different resolving powers (see text).
}
\end{figure*}

\section{Introduction}

Observations in the far red interval (7650--8750~\AA) are becoming 
increasingly used for determination of basic stellar parameters. 
The advantages of this interval are its low sensitivity to 
reddening, high photon budget for late--type stars, 
good sensitivity of modern CCD detectors, and above all richness 
of its spectral features.

The far red interval includes several important spectral lines 
(Munari 1999): ($a$\,) the K~I infrared resonant doublet 
(7664.907~\AA, 7698.979~\AA), ($b$\,) the Na~I non-resonant doublet 
(8183.256~\AA, 8194.821~\AA), ($c$\,) the O~I triplet 
(8446.247, 8446.359, 8446.758~\AA), ($d$\,) the Fe~I multiplets 60, 401 and 
others with excitation potentials between 2.17 and 5.00~eV, 
($e$\,) the head of the Paschen series of hydrogen, and, most importantly, 
($f$\,) the Ca~II non-resonant infrared triplet (8498.018, 8542,089, 
8662.140~\AA). The latter is present in all spectral types between 
B8 and M. A further advantage is that the interval of 8400 to 8750~\AA\  
is nearly free from telluric absorptions (Munari 1999).
 
The Ca~II triplet is a powerful diagnostic tool: the strength of Ca~II 
lines depends on metallicity, but it is essentially insensitive to age of 
stellar population (Garcia-Vargas, Molla \&\ Bressan 1998;
Schiavon, Barbuy \&\ Bruzual 2000; Vazdekis et al.\ 2003). These 
results have been used to study behaviour of Ca~II in composite stellar 
populations and to attempt to disentangle their age and metallicity. Composite 
systems include normal elliptical galaxies 
(Molla \&\ Garcia-Vargas 2000; Saglia et al.\ 2002; 
Cenarro et al.\ 2003), dwarf ellipticals (Michielsen, De Rijcke \&\ 
Dejonghe 2003) and active galaxies (van Groningen 1993; Diaz, Terlevich \&\ 
Terlevich 1989;  Nelson \&\ Whittle 1999; 
Schinnerer, Eckart \&\ Tacconi 2001; Marquez et al.\ 2003).
Cenarro et al.\ (2002) studied the behaviour of Ca~II line-strength 
indices in terms of effective temperature, surface gravity, and 
metallicity. Their analysis was based on a set of 706 real spectra. 
A much larger synthetic spectral library presented in this paper 
can be used to complement these results on Ca~II and other lines 
in the red domain.

Two of the largest forthcoming spectral surveys are centered on the 
far red spectral interval. GAIA is the cornerstone 6 mission of ESA, 
approved for a launch around 2010. It is aimed at providing micro-arsec 
astrometry and $\sim 10$-band photometry for $\sim 10^9$ stars brighter 
than $V=20$. Brighter targets ($V<17.5$) will also be observed by an 
on-board spectroscopic instrument operating in the 8480--8747~\AA\ 
wavelength range at a resolving power 
$R \equiv \lambda / \Delta \lambda = 11\,500$ (Katz 2003). 
The other survey, Radial Velocity Experiment (RAVE), has just started 
in April (Steinmetz 2003). This is an international collaboration which uses 
the UK-Schmidt telescope at the Anglo Australian Observatory (AAO) equipped with 
a fiber-optic spectrograph to obtain spectra in a similar wavelength domain 
(8410--8750~\AA) and at a resolving power $R \sim 8\,500$. The goal is 
to observe 35~million stars brighter than $V=16$ at declinations suitable 
for AAO. The primary motivation for both surveys is to obtain stellar 
radial velocities to be used in studies of Galactic kinematics. 
But a cross-correlation with a library of stellar spectra with known 
values of physical parameters can 
yield much more than radial velocity: in fact the effective temperature, 
gravity, metallicity (with certain element abundances), rotational 
velocity, and the presence of different kinds of peculiarities can 
all be determined. Therefore any software development for 
the GAIA mission as well as analysis of the first RAVE data needs to
use an extensive library of stellar spectra in the far red domain. The same holds 
for individual spectroscopic studies in this wavelength range which 
are becoming ever more numerous owing to the preparation of the GAIA 
mission (see e.g.\ Munari et al.\ 2001, Zwitter et al.\ 2003, 
Marrese et al.\ 2004). 

\begin{figure}[ht]
\centerline{\psfig{file=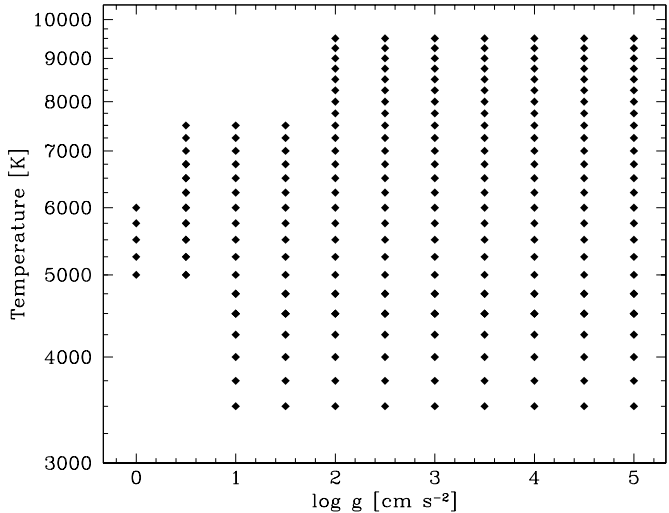,angle=270,width=10.0cm}}
\caption[]{As Fig.~1, but for models with alpha-enhancement 
([$\alpha / $Fe]$ = + 0.4$). All models were calculated for the 
metallicities $\mathrm{[M/H]} =$ --0.5, --1.0, --1.5, and --2.0.
}
\end{figure}

\begin{figure}[ht]
\centerline{\psfig{file=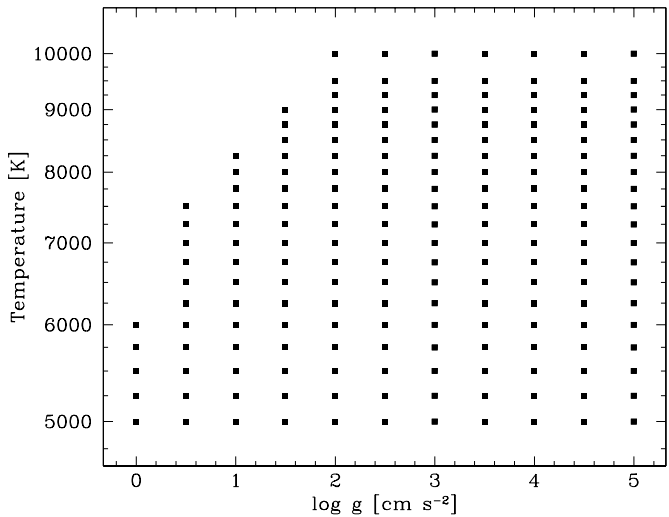,angle=270,width=10.0cm}}
\caption[]{As Fig.~1, but for models with micro-turbulent velocities 
of 0 and 4 km~s$^{-1}$. All models have solar abundances. 
}
\end{figure}

\begin{table}
\caption{Properties of models at different metallicities. The first three 
columns give values of metallicity, $\alpha$-enhancement coefficient and 
microturbulent velocity ($\xi$). The last two columns state if the 
"no-overshooting" approximation to the convection treatment and if new 
opacity distribution functions were used. }
\begin{tabular}{rcccc} \hline \hline
$[\mathrm{M}/\mathrm{H}]$ & [$\alpha$/Fe] & $\xi$ [km~s$^{-1}$] & NOVER &ODFNEW \\ \hline
$+$0.5 & 0.0 &  2 & yes& no \\ 
$+$0.2 & 0.0 &  2 &  no& no \\ 
   0.0 & 0.0 &  0 & yes& no \\ 
   0.0 & 0.0 &  2 & yes& no \\ 
   0.0 & 0.0 &  4 & yes& no \\ 
 --0.2 & 0.0 &  2 & no & no \\ 
 --0.5 & 0.0 &  2 & yes& no \\ 
 --0.5 & 0.4 &  2 & yes& yes\\ 
 --1.0 & 0.0 &  2 & yes& no \\ 
 --1.0 & 0.4 &  2 & yes& yes\\ 
 --1.5 & 0.0 &  2 & yes& no \\ 
 --1.5 & 0.4 &  2 & yes& yes\\ 
 --2.0 & 0.0 &  2 & yes& no \\ 
 --2.0 & 0.4 &  2 & yes& no \\ 
 --2.5 & 0.0 &  2 & yes& no \\ 
 --3.0 & 0.0 &  2 &  no& no \\ 
 --4.0 & 0.0 &  2 &  no& no \\ 
 \hline
\end{tabular}
\end{table}

Systematic observations of MKK standard stars in the far red spectral 
domain were presented by Munari \&\ Tomasella (1999) and by Marrese, 
Boschi \&\ Munari (2003). It is difficult
however to obtain high--signal--to--noise--ratio observations to map all 
relevant combinations of stellar parameters in a uniform manner.
So a parallel effort was launched to calculate 
a grid of synthetic spectra using ATLAS9 models from Kurucz. The first 
two papers (Munari \&\ Castelli 2000, Castelli \&\ Munari 2001) 
explored the grid in the temperature-gravity-metallicity space 
for non-rotating stars, assuming a micro-turbulent 
velocity of 2 km~s$^{-1}$. Altogether 952 spectra were presented. 

These early works were aimed to coarsely but rapidly explore an
essentially unknown wavelength range to assist early planning and
instrument design for GAIA. This phase is now over, and the community
requirements are now moving toward data and analysis reduction pipelines.
A much more extended and complete grid is now essential, and to
provide one is the aim of this paper. The grid made available with this
paper should meet the community requirements in testing algorithm coding
for some time to come. The one to be used to  analyze the actual GAIA data
when they will be finally assembled around 2015--2018 will be computed
only in the next decade taking full advantages of the
continuous advacements in the input physics, coding and atomic constants
that will be reached by that time. The present grid is also timely
presented to assist with analysis of RAVE spectra that are already
routinely obtained at AAO.

The present grid extends earlier calculations by adding more spectra, 
inclusion of stellar rotation and presentation of results 
at different spectral resolving powers. Furthermore a limited number of 
spectra corresponding to enhanced $\alpha$-element abundances and 
different values of micro-turbulent velocity are presented. The database 
consists of more than 183\,500 spectra and is freely 
available in electronic form via CDS as well as via the ESA webserver.

\section{Computation of synthetic spectra}

We computed synthetic spectra for the 7650-8750 A interval for almost all 
ATLAS9 grids of model atmospheres with different  metallicities and 
micro-turbulent velocities available at the Kurucz web-site 
(http://kurucz.harvard.edu). We generally used the NOVER models which 
differ from the previous Kurucz (1993a) models for the convection in 
that they were recomputed by Castelli with the overshooting option for
the convection switched off (Castelli et al. 1997). When NOVER grids 
were not available, we used the Kurucz (1993a) models.
Because new ATLAS9 models based on updated Opacity Distribution Functions
(ODFNEW models) have been recently computed 
(Castelli \& Kurucz 2003), we used them for a few metallicities
 with $T_\mathrm{eff} \ge 5000$~K and 
for all the spectra with $T_\mathrm{eff} \le 4750$~K.
In fact, the main difference between the NOVER models and the ODFNEW 
models is the use of new Opacity Distribution Functions (ODFs) 
computed with TiO lines from Schwenke (1998) instead of from 
Kurucz, and including H$_2$O lines, not considered at all in the previous 
ODFs. Also the ODFNEW models computed up to now are available at the
Kurucz web-site (ODFNEW grids). Details of the adopted models are 
given in Table 1. 

\begin{figure*}[t]
\centerline{\psfig{file=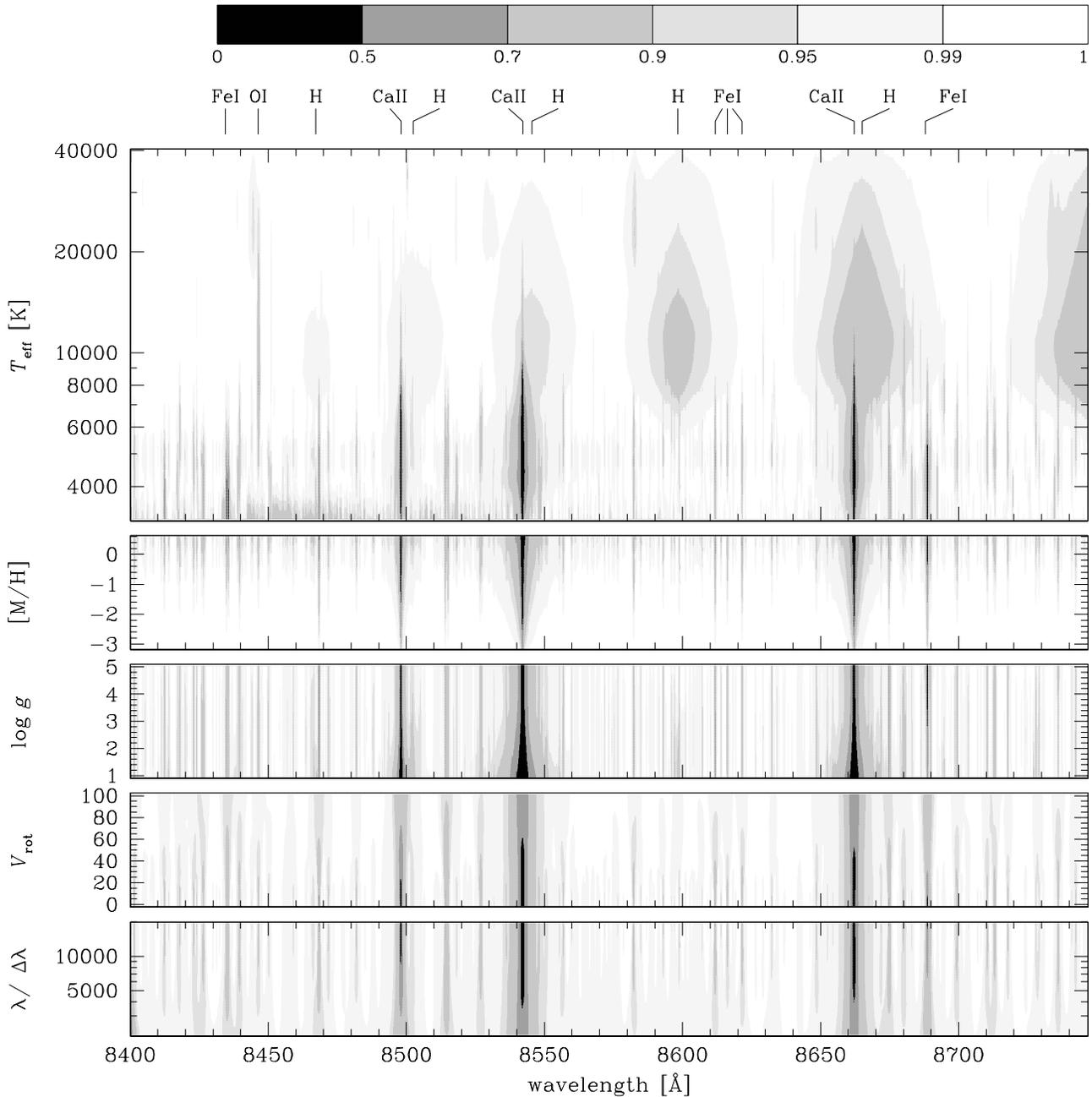,width=19cm}}
\caption[]{Cross-sections through the multidimensional 
[Temperature, Metallicity, Gravity (cm~s$^{-2}$), Rotation, 
Resolving power] data-cube of computed spectra. Parameter values 
correspond to a non-rotating K0~V type star ($T_\mathrm{eff}=5250$~K, 
[M$/$H$]=0.0$, $\log g = 4.5$, $V_\mathrm{rot}=0$~km~s$^{-1}$, 
$\xi = 2$~km~s$^{-1}$, $\lambda / \Delta \lambda = 20\,000$), except 
for the parameter which is allowed to vary on a particular plot. 
All spectra were normalized by a single cubic spline fit 
to their upper envelope. The number of greyscale hues  
(identified at the top of the figure) is intentionally kept 
small, so as to allow reading the depth of individual 
spectral lines from the graph. Ions giving rise to the most 
prominent spectral lines are identified at the top. Synthetic 
spectra span a wider wavelength range and include variation of additional 
parameters ($\alpha$-enhancement, different micro-turbulent velocities)
not shown here for clarity.
}
\end{figure*}

\begin{table*}
\caption{General ranges of parameters of the spectra. Not all combinations 
were calculated (for details see Figs.\ 1--3).}
\setlength\tabcolsep{0pt}
\begin{tabular}{lrclll} \hline \hline
&\\
Parameter &\multicolumn{3}{l}{\ \ \ \ \ Range}& Step or Values\\ \hline
&\\
Temperature & 3\,500&...&47\,500~K &step 250~K ($T_\mathrm{eff} \le 10\,000$~K), 500 or 1000~K ($T_\mathrm{eff} > 10\,000$~K)\\
Metallicity  & +0.5&...&--3.0	  &step 0.5 \&\ values of +0.2 and --0.2\\ 
Gravity     & 0.0&...&5.0	  &step 0.5 \\
$\alpha$--enhancement& 0.0&...&0.4	& (0.0, 0.4)\\
Micro-turbulence & 0&...&4  km~s$^{-1}$ & (0, 2, 4 km~s$^{-1}$)\\
Rotation velocity &0&...&100 km~s$^{-1}$ ($T_\mathrm{eff} < 7000$~K) \ \ \ \ \ \ & 
	(0, 2, 5, 10, 15, 20, 30, 40, 50, 75, 100 km~s$^{-1}$)\\
		 &0&...&500 km~s$^{-1}$ ($T_\mathrm{eff} \ge 7000$~K) & 
	(0, 10, 20, 30, 40, 50, 75, 100, 150, 200, 250, 300, 400, 500 km~s$^{-1}$)\\
Resolving power \ \ \ \ \ \ \ \ & 8\,500&...&20\,000	&(8\,500, 11\,500, 20\,000)\\ 
Wavelength  & 7653&...&8747 \AA & 2 pixels per resolution element (2.5 pixels at $R=20\,000$)\\ \hline
\end{tabular}
\setlength\tabcolsep{6pt}
\end{table*}

The synthetic spectra were computed with the SYNTHE code of Kurucz (1993b)
at a resolving power of 500\,000. The source of atomic data was Kurucz 
\&\ Bell (1995), the source of molecular data, except TiO,  was Kurucz 
(1993c), while the source of TiO data was Kurucz (1999a), who supplied 
Schwenke's (1998) computations in SYNTHE format.  Di-atomic 
molecules used in model computation are discussed in Kurucz (1992) and  
Kurucz (1993d). The source for H$_2$O data is Partridge \&\ Schwenke (1997),
as distributed by Kurucz (1999b). The solar abundances are from 
Anders \&\ Grevesse (1989), 
except for the spectra based on the ODFNEW models. In this case the solar 
abundances are from Grevesse \&\ Sauval (1998). Each spectrum computed 
for a given model atmosphere was then broadened for several values of 
rotational velocity $V_\mathrm{rot}$ and for three resolving powers: 
$R=$8\,500, 11\,500, and 20\,000  by assuming a Gaussian instrumental profile.
These resolutions were chosen because they correspond 
to the baselined values for the RAVE survey, the GAIA mission and a 
typical Cassegrain-fed Echelle spectrograph, respectively. The resulting 
final synthetic spectra were resampled to 2 pixels per resolution element 
($R=8\,500, 11\,500$) or 2.5 pixels per resolution element ($R=20\,000$).

\section{Grid of synthetic spectra}

Ranges and steps for all seven basic parameters of the grid of synthetic 
spectra are given in Table~2. We adopt a common convention of quoting 
metallicity and enhancement of $\alpha$--elements in logarithmic units with 
respect to the solar values. The gravity is in logarithmic cgs units. Details 
of all calculated parameter combinations are given in Figures~1--3. Spectra 
are placed in gravity--temperature planes, with metallicity coded by a symbol 
type. Figure~1 covers the most numerous spectra, i.e.\ the ones 
with no $\alpha$--enhancement and with microturbulent velocity of 
2~km~s$^{-1}$. The computed spectra cover the whole gravity--temperature 
plane except for hot low-gravity models which are not radiatively stable. 
Low--temperature spectra ($T_\mathrm{eff} < 5000$~K) were computed for a 
sparser set of metallicities due to large requirements of computing time. 
These spectra will be added online when completed. Fig.~2 corresponds to 
$\alpha$--enhanced cases and Fig.~3 to those with a different value of 
microturbulent velocity. Note that each of the symbols actually corresponds 
to 11 ($T_\mathrm{eff} < 7000$~K) or 14 ($T_\mathrm{eff} \ge 7000$~K) spectra 
with different values of rotational velocity (see Table~1) and to three 
different resolving powers. 

All spectra are available as ascii files grouped into different directories according 
to their resolving power and temperature. The filenames are in a standard format 
identified in Table~3. So  {\sl f765875v010r20000m05t05250g45k2nover.asc}
corresponds to a flux calibrated spectrum between  7650 and 8750~\AA, with 
$V_\mathrm{rot} = 10 $~km~s$^{-1}$, $\lambda / \Delta \lambda = 20\,000$, 
$[\mathrm{M} / \mathrm{H} ] = -0.5$, $T_\mathrm{eff} = 5250$~K, $\log g = 4.5$, 
$\xi = 2$~km~s$^{-1}$, and no $\alpha$--enhancement. 

The calculated grid is by far too large to present all of its properties here, 
so we explore only sample 
cross-sections across the grid. Figure~4 is a greyscale presentation of the 
spectra which were normalized to enhance line visibility. Each panel shows 
variation along one parameter axis, starting from a spectrum of a 
non-rotating K0~V type star. Note that all spectra were calculated in a 
wider wavelength domain, but only the 8400--8750~\AA\  range is plotted for 
clarity. 

The temperature panel of Fig.~4 clearly shows the importance of sharp 
Ca~II lines for any radial velocity study. The panel is a textbook example of the 
expected behaviour of the Paschen lines and metallic lines. The metallicity
panel illustrates that the Ca~II lines remain strong even at the lowest metallicities 
and the gravity panel shows their presence in all luminosity classes. The rotational 
velocity and resolving power panels show how the lines get smeared at high rotational 
velocities or if observing at low resolving powers.

The steps in the calculated grid are relatively small, but the coverage is not 
continuous. As an example, the step in temperature is 250~K 
(for $T_\mathrm{eff} \le 10\,000$~K). 
This is larger than the baselined accuracy of temperature determination for 
both GAIA and RAVE surveys. So the grid will have to be interpolated to smaller 
steps. Figure~5 illustrates the errors introduced by a simple linear interpolation. 
At a certain grid point $i$ with the parameter values $p_i$ we compare the 
true synthetic spectrum $S(p_i)$ with the spectrum $S'$ obtained from a linear 
combination of spectra at neighbouring grid points: 
$S' = f_{i-1} S(p_{i-1}) + f_{i+1} S(p_{i+1})$. The weights $f_{i-1}$ and $f_{i+1}$ 
are optimized so that $\int [S(p_i) - S']^2 d\lambda$ 
is minimal. The difference between the interpolated values of parameters $p'$ and 
the true ones $p_i$ can then be expressed in units of a grid step:
\begin{equation}
\nonumber
\Delta  \equiv \frac{p'-p_i}{p_{i+1}-p_{i-1}} = 
\frac{f_{i+1} (p_{i+1}-p_i) + f_{i-1} (p_{i-1}-p_i)}{(f_{i+1} + f_{i-1})(p_{i+1}-p_{i-1})}
\end{equation} 
Figure~5 shows that linear interpolation is accurate to $\simlt 10$~\%\ of the 
grid step. Note that this is the worst case scenario, corresponding to a reconstruction 
of the spectrum at the middle of the grid interval. Linear interpolation would be 
more accurate for spectra lying closer to one of the grid points. The results could 
be improved further by employing non-linear interpolation schemes. One may conclude 
that linear interpolation itself does not introduce errors exceeding 25~K in 
temperature (for $T_\mathrm{eff} < 10\,000$~K), 
0.05~dex in $[\mathrm{M}/\mathrm{H}]$ or $\log g$ and 1~km~s$^{-1}$ in 
$V_\mathrm{rot}$. Note that other errors are more important: degeneracy of parameter 
values fitting spectra with a limited signal to noise ratio complicates their 
determination (Bailer-Jones 2003, see also Fig.~1 in Zwitter 2002). Also, 
spectra of real stars do not correspond exactly to the synthetic spectra due to 
their peculiarities (e.g.\ emission lines, varied abundances of individual elements,
non-LTE effects, and non-static atmospheric structure).

\begin{figure}[ht]
\centerline{\psfig{file=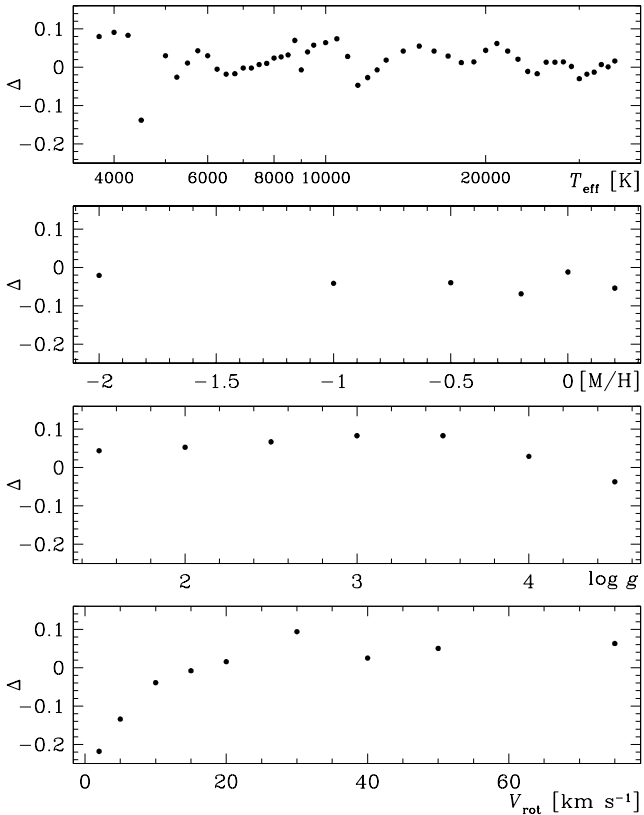,width=9.3cm,angle=0}}
\caption[]{Errors introduced by linear interpolation of the grid.
Each graph explores variation of one parameter, 
with the values of other parameters being fixed at: 
$T_\mathrm{eff} =5250$~K, [M$/$H]=0.0, $\log g = 4.5$, $V_\mathrm{rot}$=0~km~s$^{-1}$,
$\mathrm{R}= 20\,000$. The ordinate is the difference between the value 
obtained from linear grid interpolation of the given parameter and 
the true value, expressed in units of the grid step. 
Note that linear interpolation yields values 
accurate to $\simlt 10$\%\ of the grid step.
}
\end{figure}

\begin{table}
\caption{Filenames of individual spectra. Meaning of corresponding characters is 
given.}
\begin{tabular}{rll}
\hline \hline
character & meaning \\
\hline
1     &  f: fluxed spectrum (erg~s$^{-1}$~cm$^{-2}$~\AA~steradian$^{-1}$)\\
      &  n: normalized to the continuum flux \\
2--4  &  starting wavelength (nm) \\
5--7  &  ending wavelength (nm) \\
9--11 &  rotational velocity $V_\mathrm{rot}$ (km s$^{-1}$)\\
13--17&  resolving power $R$ \\
18    &  p: [M$/$H] $\ge 0.0$\\
      &  m: [M$/$H] $< 0.0$\\
19--20&  10 $\times$ ABS([M$/$H]) \\
21    &  t: no $\alpha$--enhancement \\
      &  a: $[\alpha / \mathrm{Fe} ] = 0.4 $\\ 
22--26&  effective temperature $T_\mathrm{eff}$ (K) \\
28--29&  10 $\times \log g$ \\
31    &  micro-turbulent velocity $\xi$ (km~s$^{-1}$)\\
\hline
\end{tabular}
\end{table}

\section{Discussion and Conclusions}

The grid of synthetic spectra in the 7653--8747\AA\ interval presented here is 
the most extensive so far. Altogether the grid contains 183\,588 ascii files.
They span three spectral resolutions corresponding to a typical 
Echelle spectrograph and the spectrographs of the RAVE and GAIA surveys. This 
should facilitate comparison of the results obtained with different instruments. 
Spectra with other resolving powers (R~$<20\,000$) can be easily computed from 
the grid.

\begin{figure*}[ht]
\centerline{\psfig{file=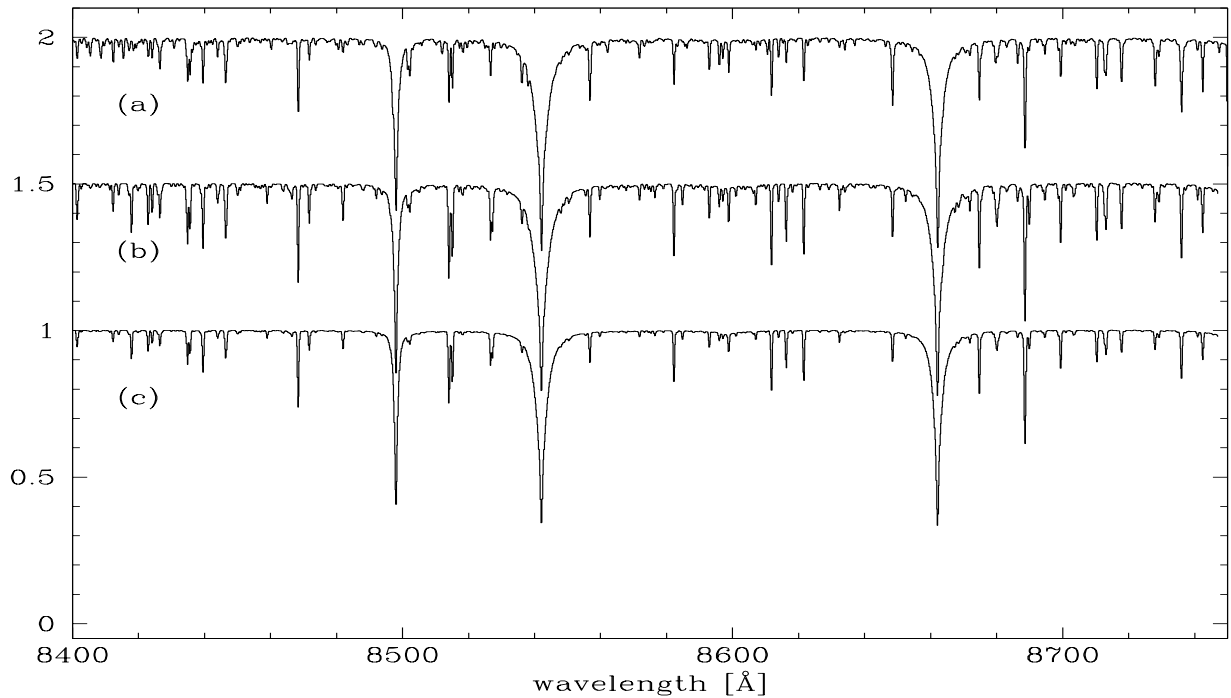,width=18.3cm,height=10.0cm,angle=270}}
\caption[]{
Comparison of computed and observed spectra: (a) observed solar flux atlas
resampled to $R=20\,000$ (Kurucz et al.\ 1984); (b) computed spectrum 
($T_\mathrm{eff} =5750$~K, [M$/$H]=0.0, $\log g = 4.5$, 
$V_\mathrm{rot}$=2~km~s$^{-1}$, $\xi = 2$~km~s$^{-1}$, 
$\mathrm{R}= 20\,000$) (c) same as (b) but for [M$/$H]=-0.5.
All spectra are plotted in normalized flux.  
The topmost two spectra have been vertically offset for clarity. 
}
\end{figure*}

The synthetic spectra we have computed can be used as templates for the 
determination of radial velocity, the primary goal of the RAVE and GAIA 
surveys. They also permit  the user to derive the primary parameters of 
stellar atmospheres: temperature, metallicity, gravity and rotational 
velocity. The grid includes subsets of spectra computed for +0.4 enhanced
abundances of the $\alpha$ elements and for different values of the
microturbulent velocity. It was shown that the grid can be easily 
interpolated to approximate spectra with parameter values between the 
steps in the grid. The errors of parameters computed with a simple linear 
interpolation do not exceed 10\%\ of the grid step. The computed spectra 
depend on the adopted values of individual element abundances. 
In particular the solar spectrum computed with Anders \&\ Grevesse's (1989) 
solar abundances  for all the elements 
(Fig.\ 6) features too strong O~I and Fe~I lines. This 
is consistent with a recently suggested lower oxygen solar abundance
$\log (N(O)/N(H))=-3.31$~dex
(Prieto, Lambert \&\ Asplund 2001) and with a lower Fe~I abundance 
$log (N(Fe)/N(H))=-4.5$~dex (Bellot Rubio \&\ Borrero 2002)
as obtained from two and three dimensional hydrodynamical model atmospheres.
This complements classical methods of diagnostic line ratios 
for the far red spectral interval discussed by Munari (2002). 

Properties of Kurucz ATLAS model atmospheres in the context of the red 
spectral interval and the GAIA mission have already been reviewed by Nesvacil 
et al.\ (2003). Plez (2003) estimated the potential of the MARCS models.
Hauschildt et al.\ (2003) presented the computations of non-stationary 
and non-spherically symmetric atmospheres with significant amounts of dust. 
Finally Th\'{e}venin, Bijaoui \&\ Katz (2003) reviewed the determination of 
chemical abundances from GAIA spectra. The common conclusion is that the 
observational potentials of the GAIA mission ask for improvements in the 
computation of synthetic stellar spectra. The RAVE collaboration is starting 
to obtain a very large number of high quality spectra in the far red 
spectral interval. They guarantee that this wavelength range is rapidly 
rising in importance. We are currently extending the grid presented here to 
a larger wavelength interval. A continuous coverage from 250~nm to 1050~nm 
should facilitate a comparison of spectra obtained in other wavelength 
domains to huge observational sets coming from the RAVE and GAIA surveys.

\acknowledgements{
We thank the anonymous referee for very relevant comments on the 
first version of the paper. 
T.Z.\ acknowledges financial support from the Slovenian Ministry for Education, 
Science and Sports, CNRS and the Royal Society as well as warm hospitality 
of GAIA groups at the Observatoire de Meudon and the Mullard Space Sciences 
Laboratory where part of this work has been completed. U.M.\ acknowledges 
financial support from the Italian Space Agency contract 
ASI-I-R-117-01 and the Italian Ministry of Education COFIN 2001 grant.
We thank R. Sordo for assisting with parts of the grid computation.
}
\\


\begin{thebibliography}{}
\bibitem{} Anders E., Grevesse N. 1989, Geochim. Cosmochim. Acta, 53, 197
\bibitem{} Bailer-Jones C.A.L. 2003, in GAIA Spectrsocopy, Science and 
  Technology (U. Munari, ed.), ASP Conf.\ Ser., 298, 199
\bibitem{} Bellot Rubio L.R., Borrero J.M. 2002, A\&A, 391, 331
\bibitem{} Castelli F., Gratton R.G., Kurucz R.L. 1997, A\&A, 318, 841
\bibitem{} Castelli F., Munari U. 2001, A\&A, 366, 1003
\bibitem{} Castelli F., Kurucz R.L. 2003, in Modelling of Stellar Atmospheres, 
  IAU Symp.\ 210 (N.E. Piskunov et al., eds.), in press
\bibitem{} Cenarro A.J., Gorgas J., Cardiel N., Vazdekis A., Peletier 
R.F. 2002, MNRAS, 329, 863
\bibitem{} Cenarro A.J., Gorgas J., Vazdekis A., Cardiel N., Peletier R.F. 
    2003, MNRAS, 339, L12
\bibitem{} Diaz A.I., Terlevich E., Terlevich R. 1989, in Active Galactic 
   Nuclei (D.E.\ Osterbrock \&\ Miller J.S., eds.), 560 
\bibitem{} Garcia-Vargas M.L., Moll\' a M., Bressan A. 1998, A\& AS,
  130, 513
\bibitem{} Grevesse N., Sauval A.J. 1998, Space Sci. Rev., 85, 161
\bibitem{} van Groningen E. 1993, A\&A, 272, 25
\bibitem{} Hauschildt P.H., Allard F., Baron E., Aufdenberg J., 
  Schweitzer A. 2003, in GAIA Spectrsocopy, Science and 
  Technology (U. Munari, ed.), ASP Conf.\ Ser., 298, 179
\bibitem{} Katz D. 2003, in GAIA Spectrsocopy, Science and 
  Technology (U. Munari, ed.), ASP Conf.\ Ser., 298, 65 
\bibitem{} Kurucz R.L., Furenlid I. Brault J., Testerman L., 1984. 
     National Solar Observatory, Sunspot, New Mexico, 
     Solar Flux Atlas from 296 to 1300nm 
\bibitem{} Kurucz R.L. 1992, RvMexAA, 23, 45
\bibitem{} Kurucz R.L. 1993a, ATLAS9 Stellar Atmospheres Programs and
  2 km s$^{-1}$ grid, CD-ROM No. 13
\bibitem{} Kurucz R.L. 1993b, SYNTHE Spectrum Synthesis Programs and 
  Line Data, CD-ROM No. 18
\bibitem{} Kurucz R.L. 1993c, Diatomic molecular data for opacity calculations,
  CD-ROM No. 15
\bibitem{} Kurucz R.L. 1993d, in Molecules in the Stellar Environment,
IAU Coll. 146 (ed. U.G. Jorgensen), 282
\bibitem{} Kurucz R.L., Bell B. 1995, Atomic Line List, CD-ROM, No. 23 
\bibitem{} Kurucz R.L. 1999a, TiO Line List, CD-ROM No. 24
\bibitem{} Kurucz R.L. 1999b, H2O Linelist from Partridge and Schwenke (1997), 
    no IDs, CD-ROM No. 26
\bibitem{} Marrese P.M., Boschi F., Munari U. 2003, A\&A, 406, 995
\bibitem{} Marrese P.M., Munari U., Siviero A., Milone E.F., 
  Zwitter T., Tomov T., Boschi F., Boeche C. 2004, A\&A, 413, 635 
\bibitem{} Marquez I., Masegosa J., Durret F., Gonzalez Delgado R.M., 
  Moles M., Maza J., Perez E., Roth M. 2003, A\&A, 409, 459
\bibitem{} Michielsen D., De Rijcke S., Dejonghe H. 2003, ApJ, 597, L21 
\bibitem{} Moll\'{a} M., Garcia-Vargas M.L. 2000, A\&A, 359, 18
\bibitem{} Munari U. 1999, Balt. A. 8, 73
\bibitem{} Munari U., Tomasella L. 1999, 137, 521
\bibitem{} Munari U., Castelli F. 2000, A\&AS, 141, 141
\bibitem{} Munari U., Tomov T., Zwitter T., Milone E.F., Kallrath J., 
Marrese P.M., Boschi F., Pr\v{s}a A., Tomasella L., Moro D. 2001, 
A\&A, 378, 477
\bibitem{} Munari U. 2002, in GAIA: A European Space Project 
  (O. Bienayme and C. Turon, eds.), EAS Publ. Ser., 2, 39
\bibitem{} Nelson C., Whittle M. 1999, Adv. Space Res., 23, 891 
\bibitem{} Nesvacil N., St\"{u}tz C., Weiss W.W. 2003, 
  in GAIA Spectrsocopy, Science and Technology (U. Munari, ed.), 
  ASP Conf.\ Ser., 298, 173 
\bibitem{} Partridge H., Schwenke D.W. 1997, J. Chem. Phys., 106, 4619
\bibitem{} Plez B. 2003, in GAIA Spectrsocopy, Science and 
  Technology (U. Munari, ed.), ASP Conf.\ Ser., 298, 189 
\bibitem{} Prieto C.A., Lambert D.L., Asplund M. 2001, ApJ, 556, L63
\bibitem{} Saglia R.P., Maraston C., Thomas D., Bender R., Colless M. 
  2002, ApJ, 579, L13
\bibitem{} Schiavon R.P., Barbuy B., Bruzual G. 2000, ApJ, 532, 453
\bibitem{} Schinnerer E., Eckart A., Tacconi L.J. 2001, ApJ, 549, 254
\bibitem{} Schwenke D.W. 1998, Chemistry and Physics of Molecules and 
  Grains in Space. Faraday Discussions No. 109. 
  The Faraday Division of the Royal Soc. of Chem., London,.321
\bibitem{} Steinmetz M. 2003, in GAIA Spectrsocopy, Science and 
  Technology (U. Munari, ed.), ASP Conf.\ Ser., 298, 381
\bibitem{} Th\'{e}venin F., Bijaoui A., Katz D.2003, in 
  GAIA Spectrsocopy, Science and Technology (U. Munari, ed.), ASP Conf.\ Ser., 
  298, 291 
\bibitem{} Vazdekis A., Cenarro A.J., Gorgas J., Cardiel N., Peletier R.F.
2003, MNRAS 340, 1317
\bibitem{} Zwitter T., 2002, A\&A, 386, 748 
\bibitem{} Zwitter T., Munari U., Marrese P.M., Pr\v{s}a A., 
  Milone E.F., Boschi F., Tomov T., Siviero A. 2003, A\&A, 404, 333 
\end{thebibliography}
\end{document}